\documentclass[reprint, superscriptaddress, amsmath, amssymb, aps, prl, floatfix, 10pt, longbibliography]{revtex4-2}

\usepackage{graphicx}
\usepackage{dcolumn}
\usepackage{array}
\usepackage{bm}
\usepackage{multirow, makecell}
\usepackage{float}

\usepackage{amsfonts}
\usepackage{amsthm}
\usepackage{siunitx}

\usepackage{verbatim}
\usepackage[dvipsnames]{xcolor}
\usepackage{hyperref}
\usepackage{amsmath}
\hypersetup{%
colorlinks = true,
urlcolor  = BrickRed,
linkcolor = NavyBlue,
citecolor = ForestGreen%
}
\begin{document}
\title{Experimental Demonstration of Quantum Pseudotelepathy}

\author{Jia-Min Xu}
\affiliation{Hefei National Research Center for Physical Sciences at the Microscale and School of Physical Sciences, University of Science and Technology of China, Hefei 230026, China}
\affiliation{CAS Center for Excellence in Quantum Information and Quantum Physics, University of Science and Technology of China, Hefei 230026, China}

\author{Yi-Zheng Zhen}
\affiliation{Hefei National Research Center for Physical Sciences at the Microscale and School of Physical Sciences, University of Science and Technology of China, Hefei 230026, China}
\affiliation{CAS Center for Excellence in Quantum Information and Quantum Physics, University of Science and Technology of China, Hefei 230026, China}

\author{Yu-Xiang Yang}
\affiliation{National Laboratory of Solid State Microstructures, School of Physics, Nanjing University, Nanjing 210093, China}
\affiliation{Collaborative Innovation Center of Advanced Microstructures, Nanjing 210093, China}

\author{Zi-Mo Cheng}
\affiliation{National Laboratory of Solid State Microstructures, School of Physics, Nanjing University, Nanjing 210093, China}
\affiliation{Collaborative Innovation Center of Advanced Microstructures, Nanjing 210093, China}

\author{Zhi-Cheng Ren}
\affiliation{National Laboratory of Solid State Microstructures, School of Physics, Nanjing University, Nanjing 210093, China}
\affiliation{Collaborative Innovation Center of Advanced Microstructures, Nanjing 210093, China}

\author{Kai Chen}
\email{Corresponding author.\\kaichen@ustc.edu.cn}
\affiliation{Hefei National Research Center for Physical Sciences at the Microscale and School of Physical Sciences, University of Science and Technology of China, Hefei 230026, China}
\affiliation{CAS Center for Excellence in Quantum Information and Quantum Physics, University of Science and Technology of China, Hefei 230026, China}

\author{Xi-Lin Wang}
\email{Corresponding author.\\xilinwang@nju.edu.cn}
\affiliation{National Laboratory of Solid State Microstructures, School of Physics, Nanjing University, Nanjing 210093, China}
\affiliation{Collaborative Innovation Center of Advanced Microstructures, Nanjing 210093, China}

\author{Hui-Tian Wang}
\email{Corresponding author.\\htwang@nju.edu.cn}
\affiliation{National Laboratory of Solid State Microstructures, School of Physics, Nanjing University, Nanjing 210093, China}
\affiliation{Collaborative Innovation Center of Advanced Microstructures, Nanjing 210093, China}

\begin{abstract}
Quantum pseudotelepathy is a strong form of nonlocality.
Different from the conventional nonlocal games where quantum strategies win statistically, e.g., the Clauser-Horne-Shimony-Holt game, quantum pseudotelepathy in principle allows quantum players to with probability 1.
In this work, we report a faithful experimental demonstration of quantum pseudotelepathy via playing the nonlocal version of Mermin-Peres magic square game, where Alice and Bob cooperatively fill in a $3\times3$ magic square.
We adopt the hyperentanglement scheme and prepare photon pairs entangled in both the polarization and the orbital angular momentum degrees of freedom, such that the experiment is carried out in a resource-efficient manner.
Under the locality and fair-sampling assumption, our results show that quantum players can simultaneously win all the queries over any classical strategy.

\end{abstract}

\maketitle
%%%%%%%%%%%%%%%%%%%%%%%%%%%%%%%%
%   Introduction  %
%%%%%%%%%%%%%%%%%%%%%%%%%%%%%%%%

Quantum advantages have been found in various areas of quantum information processing, from communication and computation to materials and engineering~\cite{aspuru-guzik_simulated_2005, liu_device-independent_2018, collaborators_hartree-fock_2020,  zhong_quantum_2020}.
Such advantages come from quantum resources, which could be entanglement, nonlocality, indistinguishability, and so on~\cite{bell_speakable_1989, mermin_hidden_1993, chitambar_quantum_2019}.
Quantum games have been widely used to reveal such quantum resources in an operational manner~\cite{kochen_problem_1967, clauser_proposed_1969, greenberger_going_1989, mermin_quantum_1990, mermin_simple_1990, peres_incompatible_1990, peres_two_1991, eisert_quantum_1999, brassard_cost_1999, brassard_quantum_2005, wilde_quantum_2017}: the players equipped with a certain quantum resource can achieve better performance than those with classical ones.
To better understand and use quantum advantages, it is important to design and develop experiments to realize such quantum games.

Quantum pseudotelepathy~\cite{brassard_quantum_2005} is a special class of quantum games: it allows quantum players to win the game with probability 1.
This distinguishes itself from conventional quantum games, where quantum players win the game statistically.
For instance, in the Clauser-Horne-Shimony-Holt (CHSH) game~\cite{clauser_proposed_1969}, quantum players win the game best with a probability approximately $0.85$ on average~\cite{brunner_bell_2014}.
There have been several types of pseudotelepathy games, including the Mermin-Greenberger-Horne-Zeilinger game~\cite{greenberger_going_1989, mermin_quantum_1990, pan_experimental_2000}, the Deutsch-Jozsa game~\cite{brassard_cost_1999}, and arguably one of the most interesting nonlocal games, the Mermin-Peres game~\cite{mermin_simple_1990, peres_incompatible_1990}.

The Mermin-Peres game is of great interest due to its simplicity and significance in quantum foundation.
In this game, the players are requested to fill in a $3\times3$ magic square with certain constraints (see detailed description below).
It can be played by a single player via sequential filling or by two separated players via cooperative filling.
The single-player version of the game reveals the contextuality of quantum theory~\cite{kochen_problem_1967, peres_two_1991,  cabello_experimentally_2008, budroni_quantum_2021} and has been carried out in various platforms~\cite{huang_experimental_2003, kirchmair_state-independent_2009, hu_experimental_2016, jerger_contextuality_2016,
dikme_measuring_2020, budroni_quantum_2021, qu_state_2021}.
The two-player version of the game, which is the focus of this work, can be used to detect nonlocality.
Particularly, the game has also been proven as the simplest for showing quantum pseudotelepathy, in the sense that it involves only two players and the number of required referee’s queries is minimized~\cite{gisin_pseudo-telepathy_2007}.
We note that the deep connection between nonlocality and contextuality has been explored, see, e.g.~\cite{abramsky_sheaf-theoretic_2011, kurzynski_fundamental_2014, zhan_contextuality_2016, dilley_more_2018, cabello_converting_2020, cabello_bell_2021, zhu_less_2021}.

In the aspect of nonlocality experiments, the loophole-free Bell tests have been carried out using the single-copy two-qubit states~\cite{hensen_loophole-free_2015, giustina_significant-loophole-free_2015, shalm_strong_2015, rosenfeld_event-ready_2017}.
Unfortunately, a faithful realization of the nonlocal Mermin-Peres game is still experimentally challenging and consuming, as it requires one to distribute {\em two} two-qubit entangled states and to carry out local measurements on two qubits from different entangled pairs.
Meanwhile, the game is in principle designed to be won with unit probability, which can never happen in realistic experiments with errors and noises.
One is required to employ the high-fidelity state preparation and measurement in experiment to beat the winning threshold.
Recently,  a scheme based on quantum dots inside optical cavities was also proposed~\cite{bugu2020surpassing}.

In this work, we faithfully demonstrate the Mermin-Peres game, and therefore quantum pseudotelepathy, using a full-photonic setup in a resource-efficient manner.
We adopt hyperentangled photons and encode two maximally entangled states on one pair of photons using 2 degrees of freedoms (DoFs), namely, the polarization and orbital angular momentum (OAM) degree.
Such an architecture strongly reduces the experimental overheads in preparations, and more importantly increases the coincidence in detection compared with the common four-photon scheme.
Furthermore, we use quantum random numbers~\cite{nie2014practical} to select queries, which in principle guarantees intrinsic randomness.
By assuming fair sampling and locality, we conclude that the players with hyperentangled states can outperform any classical strategy to win all queries with significant high probabilities.

\begin{figure}[!htp]
\centering
\includegraphics[width=0.95\linewidth]{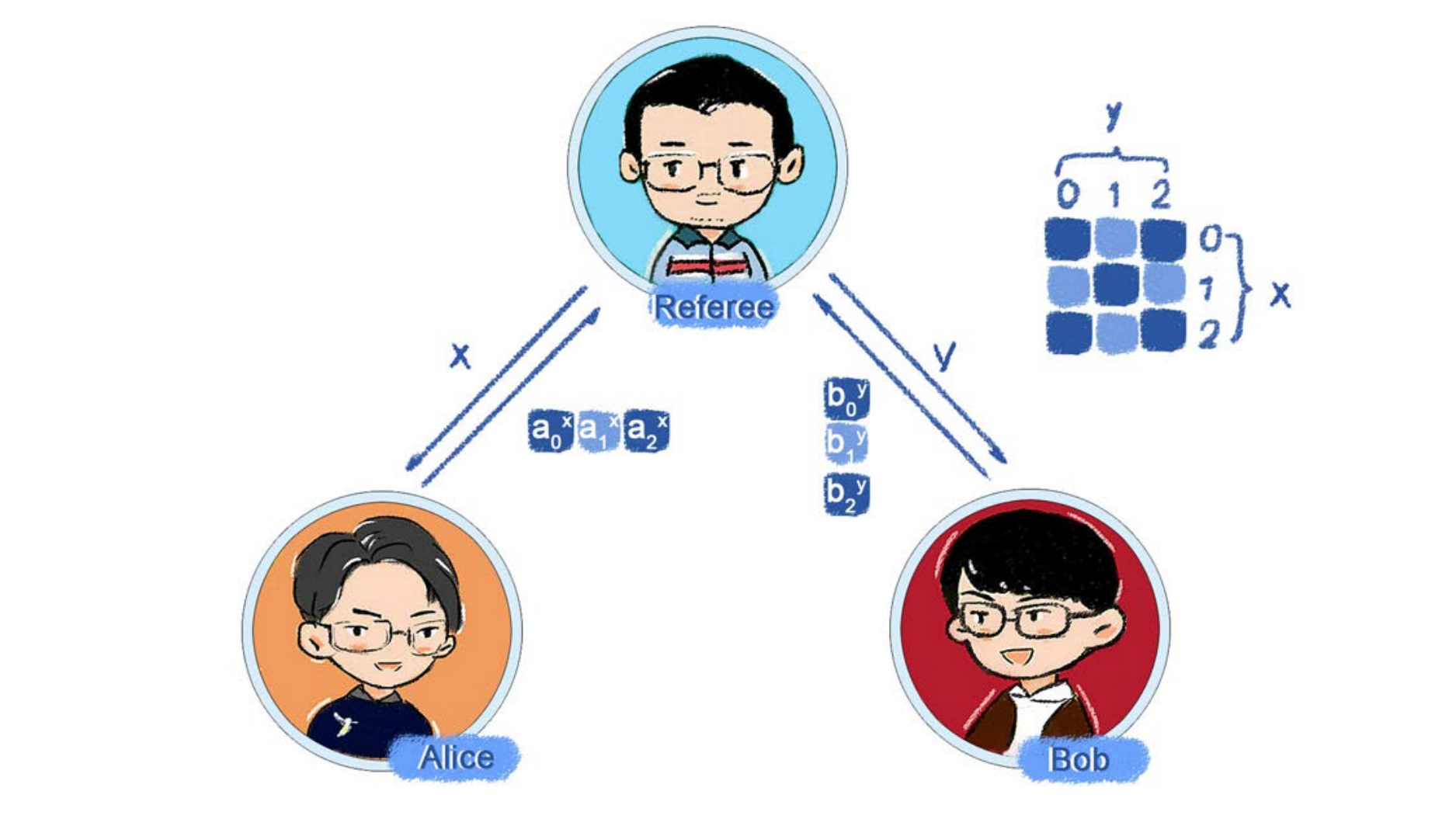}%\vspace{3mm}
\caption{\label{fig:scheme}The Mermin-Peres magic square game.
The referee randomly sends $x$ and $y$ to Alice and Bob, respectively.
Alice and Bob reply with three numbers, either +1 or -1, in a row and in a column, respectively.
They win the game if the overlapped entry is the same.
}%\vspace{5mm}

\end{figure}

\textit{Mermin-Peres game}---%
In the Mermin-Peres game, two players, Alice and Bob, are required to independently fill a $3\times 3$ magic square.
As shown in Fig.~\ref{fig:scheme}, the referee randomly sends two queries $x,y\in\{0,1,2\}$ to Alice and Bob, respectively.
Here, $x$ labels rows and $y$ labels columns.
Alice and Bob are required to reply with three numbers with specific conditions.
Denote Alice's answers in a row as $[a_0^x,a_1^x,a_2^x]$ and Bob's answers in a column as $[b_0^y,b_1^y,b_2^y]^T$, where $a_i^x,b_j^y\in\{-1,+1\}$ for any $i,j\in\{0,1,2\}$.
Alice's answers must satisfy $\prod_i a_i^x=+1$, while Bob's should satisfy $\prod_j b_j^y=-1$ for any $x$ and $y$.
During the game, Alice and Bob are forbidden to communicate with each other.
They win the game if the overlapped entry of Alice's row and Bob's column is always the same, i.e., $a_y^x=b_x^y$ for each $x$ and $y$.

% figure1

Classical strategies cannot always win the game.
One can prove it by contradiction~\cite{mermin_simple_1990, peres_incompatible_1990}.
Considering a deterministic classical strategy what gives definite answers when receiving queries $x$ or $y$.
Then, if one multiplies all Alice's answers, she would have $\prod_{x,i}a_i^x=+1$, while if one multiplies all Bob's answers, he would have $\prod_{y,j}b_j^y=-1$.
That is, for any deterministic classical strategy, Alice and Bob cannot successfully fill the same number for all entries in the square.
There are nine optimal deterministic strategies, as depicted in Table~\ref{tab:classstrategy}.
For each strategy, Alice and Bob win eight out of the total nine query pairs $(x,y)$.
A general classical strategy can be viewed as a mixture of deterministic strategies, so one can conclude that any classical strategy cannot simultaneously win each query pair with probabilities higher than $8/9$ \footnote{Suppose a classical strategy is a mixture of nine optimal deterministic strategies with probabilities $p_1,\dots,p_9$. Since $\sum_i p_i =1$, one must have one $p_i$ not less than $1/9$. Then, for the classical strategy, there must be a query pair where the winning probability is not larger than $8/9$.}.
We will use this criterion to verify the quantum pseudotelepathy in our experiment.
We also refer the classical strategy that wins each query pair with probability $8/9$ as the unbiased optimal classical strategy.

\begin{table}
\begin{tabular}{cc|c|c|c|}
\multirow{1}{*}{} & \multicolumn{1}{c}{} & \multicolumn{3}{c}{$\overbrace{\hspace{1cm}}^y$}\tabularnewline[-.2cm]
 & \multicolumn{1}{c}{\multirow{1}{*}{}} & \multicolumn{1}{c}{0} & \multicolumn{1}{c}{1} & \multicolumn{1}{c}{2}\tabularnewline
\cline{3-5} \cline{4-5} \cline{5-5}
\multirow{3}{*}{$x {\Bigg \{}$}\hspace{-.1cm} & 0 & 1 & 1 & 1\tabularnewline
\cline{3-5} \cline{4-5} \cline{5-5}
 & 1 & -1 & -1 & 1\tabularnewline
\cline{3-5} \cline{4-5} \cline{5-5}
 & 2 & 1 & 1 & \#\tabularnewline
\cline{3-5} \cline{4-5} \cline{5-5}
\end{tabular}
\hfill
\begin{tabular}{cc|c|c|c|}
\multirow{1}{*}{} & \multicolumn{1}{c}{} & \multicolumn{3}{c}{$\overbrace{\hspace{1cm}}^y$}\tabularnewline[-.2cm]
 & \multicolumn{1}{c}{\multirow{1}{*}{}} & \multicolumn{1}{c}{0} & \multicolumn{1}{c}{1} & \multicolumn{1}{c}{2}\tabularnewline
\cline{3-5} \cline{4-5} \cline{5-5}
\multirow{3}{*}{$x {\Bigg \{}$}\hspace{-.15cm} & 0 & 1 & 1 & 1\tabularnewline
\cline{3-5} \cline{4-5} \cline{5-5}
 & 1 & -1 & 1 & -1\tabularnewline
\cline{3-5} \cline{4-5} \cline{5-5}
 & 2 & 1 & \# & 1\tabularnewline
\cline{3-5} \cline{4-5} \cline{5-5}
\end{tabular}
\hfill
\begin{tabular}{cc|c|c|c|}
\multirow{1}{*}{} & \multicolumn{1}{c}{} & \multicolumn{3}{c}{$\overbrace{\hspace{1cm}}^y$}\tabularnewline[-.2cm]
 & \multicolumn{1}{c}{\multirow{1}{*}{}} & \multicolumn{1}{c}{0} & \multicolumn{1}{c}{1} & \multicolumn{1}{c}{2}\tabularnewline
\cline{3-5} \cline{4-5} \cline{5-5}
\multirow{3}{*}{$x {\Bigg \{}$}\hspace{-.15cm} & 0 & 1 & 1 & 1\tabularnewline
\cline{3-5} \cline{4-5} \cline{5-5}
 & 1 & 1 & -1 & -1\tabularnewline
\cline{3-5} \cline{4-5} \cline{5-5}
 & 2 & \# & 1 & 1\tabularnewline
\cline{3-5} \cline{4-5} \cline{5-5}
\end{tabular}

\vspace{1.5mm}

\begin{tabular}{cc|c|c|c|}
\multirow{1}{*}{} & \multicolumn{1}{c}{} & \multicolumn{3}{c}{$\overbrace{\hspace{1cm}}^y$}\tabularnewline[-.2cm]
 & \multicolumn{1}{c}{\multirow{1}{*}{}} & \multicolumn{1}{c}{0} & \multicolumn{1}{c}{1} & \multicolumn{1}{c}{2}\tabularnewline
\cline{3-5} \cline{4-5} \cline{5-5}
\multirow{3}{*}{$x {\Bigg \{}$}\hspace{-.15cm} & 0 & 1 & 1 & 1\tabularnewline
\cline{3-5} \cline{4-5} \cline{5-5}
 & 1 & 1 & 1 & \#\tabularnewline
\cline{3-5} \cline{4-5} \cline{5-5}
 & 2 & -1 & -1 & 1\tabularnewline
\cline{3-5} \cline{4-5} \cline{5-5}
\end{tabular}
\hfill
\begin{tabular}{cc|c|c|c|}
\multirow{1}{*}{} & \multicolumn{1}{c}{} & \multicolumn{3}{c}{$\overbrace{\hspace{1cm}}^y$}\tabularnewline[-.2cm]
 & \multicolumn{1}{c}{\multirow{1}{*}{}} & \multicolumn{1}{c}{0} & \multicolumn{1}{c}{1} & \multicolumn{1}{c}{2}\tabularnewline
\cline{3-5} \cline{4-5} \cline{5-5}
\multirow{3}{*}{$x {\Bigg \{}$}\hspace{-.15cm} & 0 & 1 & 1 & 1\tabularnewline
\cline{3-5} \cline{4-5} \cline{5-5}
 & 1 & 1 & \# & 1\tabularnewline
\cline{3-5} \cline{4-5} \cline{5-5}
 & 2 & -1 & 1 & -1\tabularnewline
\cline{3-5} \cline{4-5} \cline{5-5}
\end{tabular}
\hfill
\begin{tabular}{cc|c|c|c|}
\multirow{1}{*}{} & \multicolumn{1}{c}{} & \multicolumn{3}{c}{$\overbrace{\hspace{1cm}}^y$}\tabularnewline[-.2cm]
 & \multicolumn{1}{c}{\multirow{1}{*}{}} & \multicolumn{1}{c}{0} & \multicolumn{1}{c}{1} & \multicolumn{1}{c}{2}\tabularnewline
\cline{3-5} \cline{4-5} \cline{5-5}
\multirow{3}{*}{$x {\Bigg \{}$}\hspace{-.15cm} & 0 & 1 & 1 & 1\tabularnewline
\cline{3-5} \cline{4-5} \cline{5-5}
 & 1 & \# & 1 & 1\tabularnewline
\cline{3-5} \cline{4-5} \cline{5-5}
 & 2 & 1 & -1 & -1\tabularnewline
\cline{3-5} \cline{4-5} \cline{5-5}
\end{tabular}

\vspace{1.5mm}

\begin{tabular}{cc|c|c|c|}
\multirow{1}{*}{} & \multicolumn{1}{c}{} & \multicolumn{3}{c}{$\overbrace{\hspace{1cm}}^y$}\tabularnewline[-.2cm]
 & \multicolumn{1}{c}{\multirow{1}{*}{}} & \multicolumn{1}{c}{0} & \multicolumn{1}{c}{1} & \multicolumn{1}{c}{2}\tabularnewline
\cline{3-5} \cline{4-5} \cline{5-5}
\multirow{3}{*}{$x {\Bigg \{}$}\hspace{-.15cm} & 0 & 1 & 1 & \#\tabularnewline
\cline{3-5} \cline{4-5} \cline{5-5}
 & 1 & 1 & 1 & 1\tabularnewline
\cline{3-5} \cline{4-5} \cline{5-5}
 & 2 & -1 & -1 & 1\tabularnewline
\cline{3-5} \cline{4-5} \cline{5-5}
\end{tabular}
\hfill
\begin{tabular}{cc|c|c|c|}
\multirow{1}{*}{} & \multicolumn{1}{c}{} & \multicolumn{3}{c}{$\overbrace{\hspace{1cm}}^y$}\tabularnewline[-.2cm]
 & \multicolumn{1}{c}{\multirow{1}{*}{}} & \multicolumn{1}{c}{0} & \multicolumn{1}{c}{1} & \multicolumn{1}{c}{2}\tabularnewline
\cline{3-5} \cline{4-5} \cline{5-5}
\multirow{3}{*}{$x {\Bigg \{}$}\hspace{-.15cm} & 0 & 1 & \# & 1\tabularnewline
\cline{3-5} \cline{4-5} \cline{5-5}
 & 1 & 1 & 1 & 1\tabularnewline
\cline{3-5} \cline{4-5} \cline{5-5}
 & 2 & -1 & 1 & -1\tabularnewline
\cline{3-5} \cline{4-5} \cline{5-5}
\end{tabular}
\hfill
\begin{tabular}{cc|c|c|c|}
\multirow{1}{*}{} & \multicolumn{1}{c}{} & \multicolumn{3}{c}{$\overbrace{\hspace{1cm}}^y$}\tabularnewline[-.2cm]
& \multicolumn{1}{c}{\multirow{1}{*}{}} & \multicolumn{1}{c}{0} & \multicolumn{1}{c}{1} & \multicolumn{1}{c}{2}\tabularnewline
\cline{3-5} \cline{4-5} \cline{5-5}
\multirow{3}{*}{$x {\Bigg \{}$}\hspace{-.15cm} & 0 & \# & 1 & 1\tabularnewline
\cline{3-5} \cline{4-5} \cline{5-5}
 & 1 & 1 & 1 & 1\tabularnewline
\cline{3-5} \cline{4-5} \cline{5-5}
 & 2 & 1 & -1 & -1\tabularnewline
\cline{3-5} \cline{4-5} \cline{5-5}
\end{tabular}
\caption{\label{tab:classstrategy}%
{\em Nine deterministic optimal classical strategies}. Here, \#$=+1$ for Alice and \#$=-1$ for Bob.
When receiving queries $x$ and $y$, Alice and Bob select one table via preshared randomness and reply with the $x$th row and $y$th column, respectively.
If they uniformly select tables, they would have on average a winning probability $8/9$ for each query.}
\end{table}

\vspace{2mm}

\begin{table}
\begin{tabular}{cc|c|c|c|}
\multirow{1}{*}{} & \multicolumn{1}{c}{} & \multicolumn{3}{c}{$\overbrace{\hspace{3.2cm}}^y$}\tabularnewline[-.2cm]
& \multicolumn{1}{c}{\multirow{1}{*}{}} & \multicolumn{1}{c}{0} & \multicolumn{1}{c}{1} & \multicolumn{1}{c}{2}\tabularnewline
\cline{3-5} \cline{4-5} \cline{5-5}
\multirow{9}{*}{$x\left\{\rule{0pt}{40pt}\right.$}\hspace{-.2cm} &&&&\tabularnewline
 & 0 & $I\otimes Z$ & $Z\otimes I$ & $Z\otimes Z$\tabularnewline
&&&&\tabularnewline
\cline{3-5} \cline{4-5} \cline{5-5}
&&&&\tabularnewline
& 1 & $X\otimes I$ & $I\otimes X$ & $X\otimes X
 $\tabularnewline
&&&&\tabularnewline
\cline{3-5} \cline{4-5} \cline{5-5}
&&&&\tabularnewline
 & 2 & $-X\otimes Z$ & $-Z\otimes X$ & $Y\otimes Y$\tabularnewline
&&&&\tabularnewline
\cline{3-5} \cline{4-5} \cline{5-5}
\end{tabular}
\caption{\label{tab:quantstrategy}%
{\em Optimal quantum strategy}. The $X,Y,Z$ are three Pauli matrices.
When receiving queries $x$ and $y$, Alice and Bob select the $x$th row ($y$th column) of observables to measure their systems.
They win all queries with probability 1.}
\end{table}

Quantum strategies can successfully fill every entry of the square with probability 1.
Suppose that in each round two maximally entangled states $\left|\Psi\right\rangle _{A_{1}A_{2}B_{1}B_{2}}=\left|\psi\right\rangle _{A_{1}B_{1}}\otimes\left|\psi\right\rangle _{A_{2}B_{2}}$ with $\left|\psi\right\rangle =\left(\left|00\right\rangle +\left|11\right\rangle \right)/\sqrt{2}$ are distributed to Alice and Bob, where Alice has systems $A_{1}A_{2}$ and Bob has $B_{1}B_{2}$.
After the referee sends them queries $x$ and $y$, respectively, they independently perform measurements $M_{A}^{\left(x,i\right)}$ and $M_{B}^{\left(y,j\right)}$ and obtain answers $a^x_i$ and $b^y_j$.
The measurement strategy is given in Table~\ref{tab:quantstrategy}.
In this table, the measurements in each row or in each column are pairwise commutative, such that they can be jointly measured \footnote{See Supplemental Material for the measurement basis of Alice and Bob’s quantum strategies, the experimental measurement scheme, and the minimal detection efficiency to close the loophole}.
It can be verified that for all $x$ and $y$, the product of Alice's output is always $+1$, as $\prod_{i}M_A^{(x,i)}=I$, while the product of Bob's output is always $-1$, as $\prod_{j}M_B^{(y,j)}=-I$.
For the game, Alice and Bob will always win each entry since $\left\langle \Psi\right|M_A^{(x,y)}\otimes M_B^{(y,x)}\left|\Psi\right\rangle =1$ $\forall x,y$, i.e., they have same value $a^x_y=b^y_x$, which exhibits quantum pseudotelepathy.
As a comparison, recall that for the usual nonlocal game (e.g., the CHSH game), quantum strategies cannot win each query with probability 1.
In this sense, quantum pseudotelepathy is a stronger form of nonlocality.

\emph{Experimental implementation}---%
In an experiment, the crucial part for our experiment is the generation of hyperentangled state.
Note that, if one prepares the entangled states individually, say, two entangled photon pairs, and perform four-photon coincidence in the measurement, not only the setup can be complicated, but also the detection efficiency will significantly drop down.
Instead, we seek a hyperentangled photon pair and encode two entangled states into 2 DoFs, i.e.,  polarization and OAM.
We use $|H\rangle$, $|V\rangle$, $|r\rangle$, and $|l\rangle$ to represent the horizontal polarization, vertical polarization, right-handed OAM of $+\hbar$, and left-handed OAM of $-\hbar$, respectively.
The prepared state is in the form of
\begin{equation}
\left|\Psi\right\rangle = \left|\psi\right\rangle _{{\rm pol}}\ensuremath{\otimes\left|\psi\right\rangle _{{\rm oam}}=\frac{\left|HH\right\rangle +\left|VV\right\rangle }{\sqrt{2}}\otimes\frac{\left|rl\right\rangle +\left|lr\right\rangle }{\sqrt{2}}.}
\label{eq:exp-state}
\end{equation}

\begin{figure*}[!hbt]
\centering
\includegraphics[width=.8\linewidth]{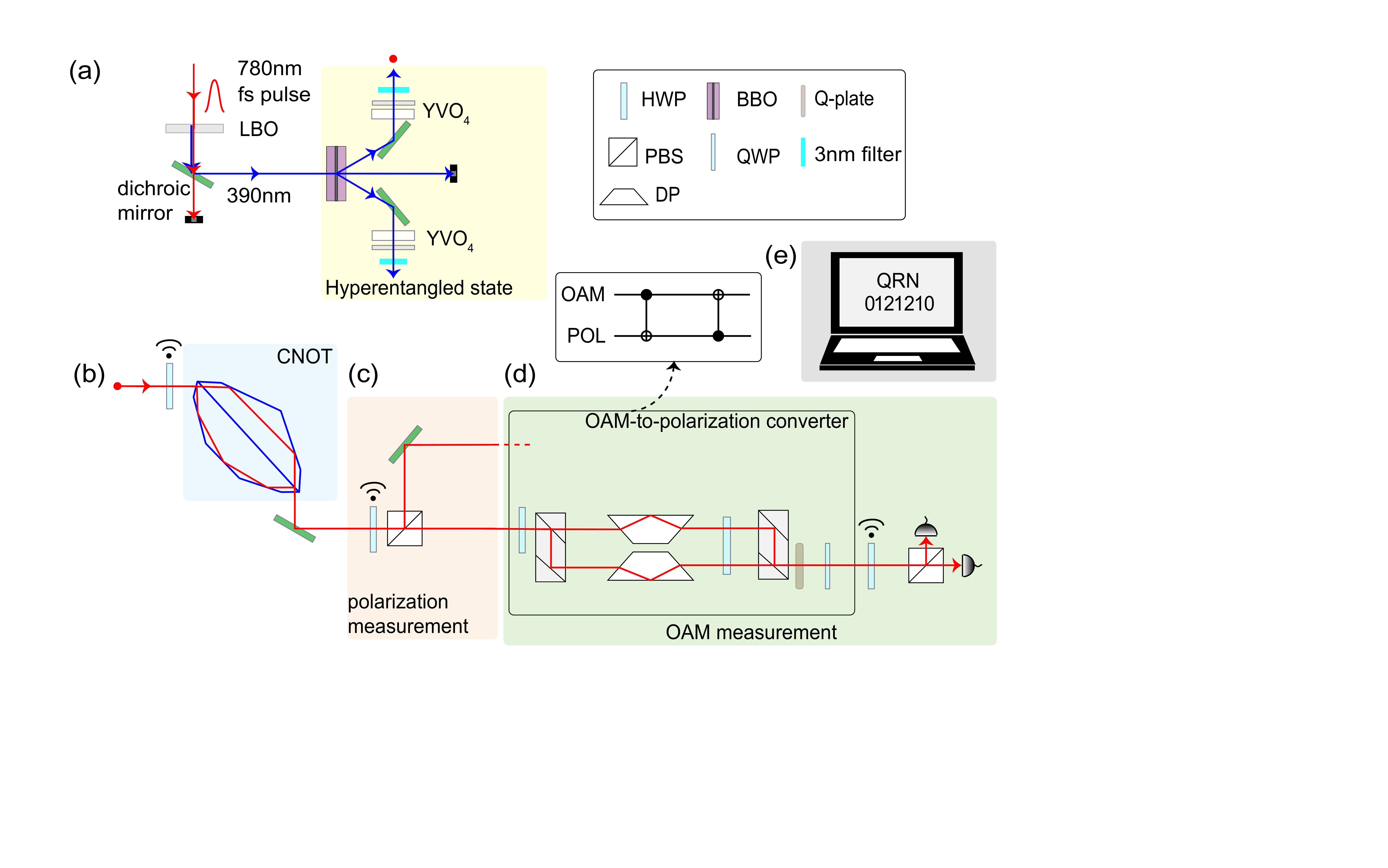}%\vspace{3mm}
\caption{Experimental setup to play quantum magic square game. (a) The generation of two-photon Polarization- and OAM hyperentangled state. (b) Polarization interferometric prism (PIP). Implement a Polarization-OAM Controlled-NOT (CNOT) gate for single photons with the polarization acting as the control qubit and the OAM as the target qubit. The HWP with a WiFi sign represents HWP that changes angle according to QRN (quantum random number). (c) Polarization measurement (d) Dual-channel OAM readout by coherently convert OAM to the polarization by a swap gate (inset). (e) Quantum random number sends $0$, $1$ and $2$ to Alice and Bob, where $0$, $1$ and $2$ denote the first, second and third row or column in the game, respectively.
OAM as the target qubit.
}
%\vspace{5mm}
\label{fig:scheme2}
\end{figure*}

To produce the hyperentangled photon pairs, two pieces of 0.6~mm thick, identically cut type-I $\beta$-barium borates (BBO) crystals with optic axes in perpendicular planes are glued together for spontaneous parametric down-conversion~\cite{barreiro2005generation, wang2015quantum} and pumped by ultrafast laser pulses with central wavelength of 390 nm, pulse duration of 150 fs, average power of 40~mW and repetition rates of 80~MHz.
The spatial and temporal walk-offs are compensated by birefringence crystals.
%The fidelity of the generated hyperentangled states are measured to be 0.928(1).
When a hyperentangled photon pair is created, one photon is sent to Alice and the other one is sent to Bob.
At the same time, the referee employs a series of quantum random numbers~\cite{nie2014practical} and distributes random numbers 0, 1, or 2 as queries to Alice and Bob.

Followed by the creation of the hyperentangled state, one needs to jointly measure the results of each of the rows and columns of Table~\ref{tab:quantstrategy}.
Notice that the observables in each of the rows or columns are mutually commuting, which means they have the common eigenstates~\cite{aravind2004quantum}.
For Alice, we choose $ZZ$ basis for the first row, $XX$ basis for the second, and graph state basis~\cite{Note2} for the last.
As for Bob, we choose $XZ$ basis for the first column, $ZX$ basis for the second, and Bell state basis~\cite{Note2} for the last.
Obtaining the expectation values of all nine observables requires $3\times3=9$ settings.
We therefore design the measurement setup to implement nine settings, as shown in Fig.~\ref{fig:scheme2}.
As a faithful demonstration of the game, the measurement settings are freely switchable.
To achieve this, we insert half-wave plates (HWPs) before the polarization interferometric prisms (PIPs), as shown in Fig.~\ref{fig:scheme2}(b),
and use stepper motors to independently drive HWPs $1\sim8$ to $0^\circ$ and $22.5^\circ$, depending on the received random number.

As an illustration of how the measurement settings are implemented, here we consider the case of $(x,y)=(0,2)$.
When random numbers $(x,y)=(0,2)$ are generated and sent to Alice and Bob, respectively, their stepper motors are triggered such that $ZZ$ is measured at Alice's side while Bell state measurement is measured at Bob's side.
For the measurement $ZZ$, a readout of qubits encoded in polarization and OAM is performed as follows.
Polarization qubit is first read out with a standard polarization analyzer, which consists of one HWP, one polarizing beam splitter (PBS).
After the PBS, the two paths, transmission and reflection of photons, correspond to two outcomes [as shown in Fig.~\ref{fig:scheme2}(c)].
While for qubits encoded in OAM, a swap gate is then used to transfer the OAM information to polarization, followed by another polarization analyzer~\cite{wang201818}.

For the measurement under Bell basis, an additional polarization-OAM CNOT gate is employed.
The CNOT gate is implemented with a PIP~\cite{ren2020}, where two prisms are glued to each other with PBS coating inside [(as shown in Fig.~\ref{fig:scheme2}(b)].
In this way, $|H\rangle$ and $|V\rangle$ components of the input photon are split into different prisms, and then recombined after different-parity number reflections.
The topological charge of an OAM state is remained after an even number of reflections and inverted after an odd number of reflections.
Thus, at the output of PIP, $X$ or $I$ can be performed on the OAM qubit according to whether the polarization is $|V\rangle$ or not, such that the  CNOT gate is realized . (Please refer to Supplemental Material~\cite{Note2} for details about other measurements.)

%%%%%%%%%%%%%%%%%%%%%%%%%%%%%%%%
%\vspace{5mm}
\emph{Results}---%
To ensure that the hyperentangled state in Eq.~(\ref{eq:exp-state}) is successfully generated in the experiment, we measure the fidelity of the prepared state.
The density matrix $\rho=\left|\Psi\right\rangle\left\langle\Psi\right|$ can be expressed as
\begin{equation}\label{eq:density}
\rho=\frac{1}{16}(II+XX-YY+ZZ)\otimes(II+XX+YY-ZZ),
\end{equation}
where $X,Y,Z$ are three Pauli operators and $I$ is the identity matrix.
We thus measure the expectation values of all involved nine observables, i.e., $XXXX$, $XXYY$, $\dots$, $ZZYY$, and $ZZZZ$.
This is done by reading out all the 16-outcome combinations of each four-qubit observable.
The fidelity of our prepared state is $F=0.928(1)$ with respect to the ideal state in Eq.~\eqref{eq:density}~\cite{Note2}.

\begin{figure}[!htp]
\centering
\includegraphics[width=1\linewidth]{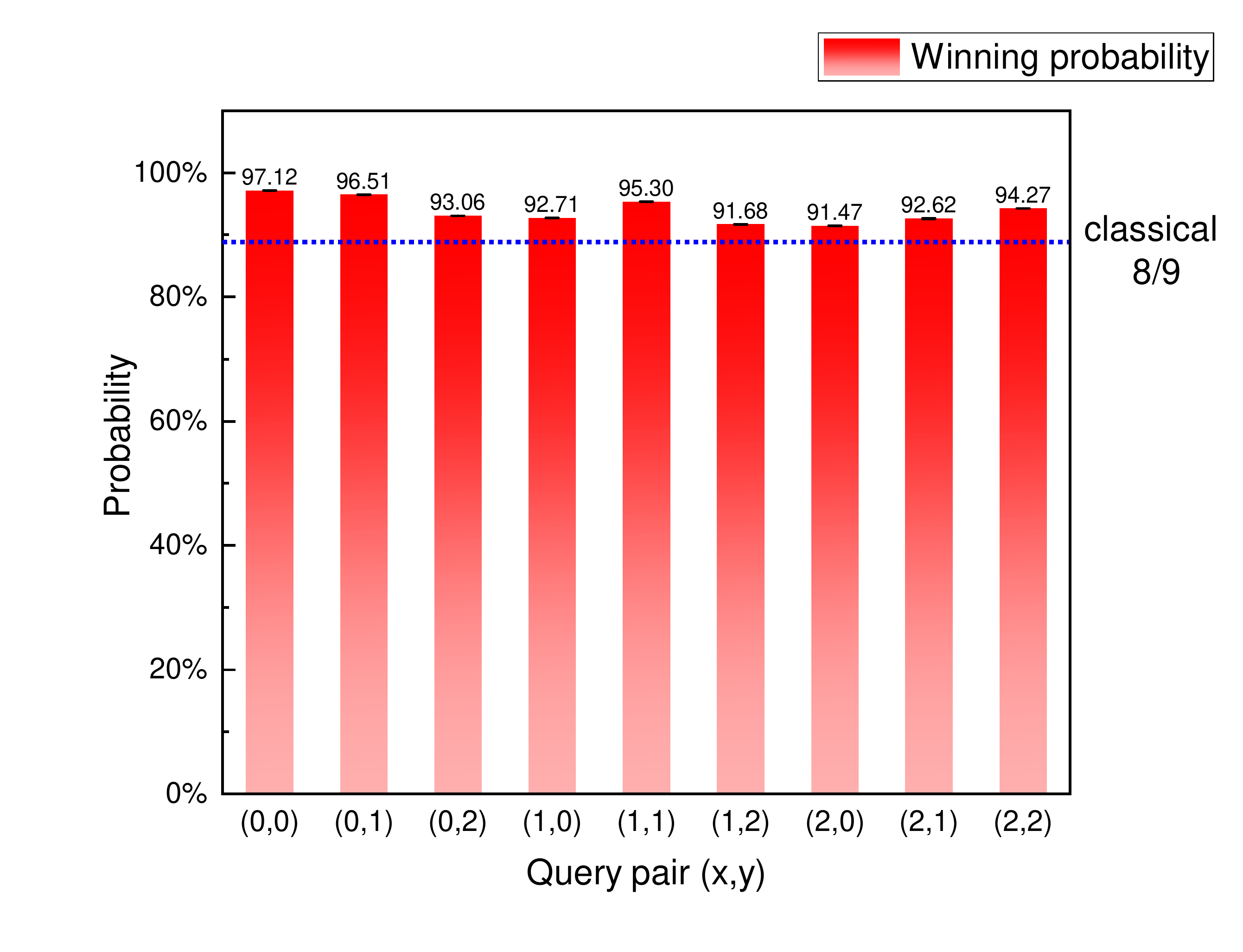}%\vspace{3mm}
\caption{\label{fig:fig3_result}Summary of results.
The winning probability for each query pair $(x,y)$ is calculated from experimental data.
The standard deviation is less than $0.01\%$, which is negligible in the histogram.
The dotted blue line represents the unbiased classical bound $8/9$.
}%\vspace{5mm}
\end{figure}

In the experiment, every coincidence event represents a round of the game.
As a result, we totally play 1\,075\,930 rounds of the game during which 1\,009\,610 rounds are won.
The average winning probability is $0.9384 (2)$, which is more than 247 standard deviations.
The precise winning probability for each query pair $(x,y)$ is shown in Fig.~\ref{fig:fig3_result}.
Compared to classical bound $8/9$ ($\approx 88.89\%$) for each query pair, we observe that the quantum theory enables players to win all the cases with probabilities over $91\%$.
This implies that the quantum strategies can outperform any classical strategy (e.g., the unbiased classical optimal strategy).
Because of the noise from the multiphoton emission of parametric down-conversion and the imperfection of the optical interferometers, the ideal 100\% winning probabilities cannot be achieved in a real experiment.
Nevertheless, one can conclude that no classical strategy can outperform the experimental result in Fig.~\ref{fig:fig3_result}, since any mixture of deterministic classical strategies (as shown in Table~\ref{tab:classstrategy}) win all nine queries with probabilities higher than $8/9$  simultaneously.
We remark that the average winning probability is slightly larger than the fidelity of the hyperentangled state because the polarization entangled qubits and the OAM entangled qubits involve different types of noises.

\emph{Conclusion}---%
Quantum pseudotelepathy has been shown to have applications for quantum communication complexity.
We experimentally demonstrated the quantum pseudotelepathy by playing the Mermin-Peres game.
We adopted multi-DoF encoding of hyperentangled photon pairs to reduce the experimental overheads, and prepared hyperentangled state with high fidelity.
Quantum random numbers are exploited to generate the referee's queries.
The experiment shows that, under an assumption of locality and fair sampling, the quantum players can fill the magic square more successfully than any classical strategy.
To close all loopholes in this experiment, not only Alice and Bob should satisfy spacelike separation during the game, but also the overall detection efficiency should be higher than 87.5\%~\cite{Note2}, which is much more difficult than the loophole-free Bell test.
Additionally, the Mermin-Peres game is extensively used as a certification of quantumness of devices in recent research.
It is an important open question to find many practically useful tasks for this game.
Our result will explicitly stimulate the design and investigation of novel protocols based on known quantum games,
especially for quantum communication~\cite{basak_2019_device} and randomness expansion~\cite{adamson_2020_quantum}.
%\vspace{5mm}

%\iffalse
\begin{acknowledgments}
We gratefully acknowledge enlightening discussions with Nai-Le Liu, Li Li, Liuwu Wang, and Francesco Buscemi when starting this work,
and thank Xin-Yu Xu, Yihan Luo, Yingqiu Mao for valuable discussion.
We are also grateful to the anonymous referees for their very constructive comments about improvements of this work.
This work was supported by
the National Natural Science Foundation of China (Grants No.~%
62031024,%kc
11575174,%kc
11922406,%xw
12005091%yz
),
National Key R\&D Program of China (2019YFA0308700),
the Chinese Academy of Sciences,
the Anhui Initiative in Quantum Information Technologies,
and
the China Postdoctoral Science Foundation (No.~2020M671856).

 J.-M. X. and Y.-Z. Z. contributed equally to this work.
\end{acknowledgments}
%\fi

\bibliography{Ref_MagicSquare}

\end{document}

\title{Supplemental Material:\\
Experimental Demonstration of Quantum Pseudotelepathy}

\author{Jia-Min Xu}
\affiliation{Hefei National Research Center for Physical Sciences at the Microscale and School of Physical Sciences, University of Science and Technology of China, Hefei 230026, China}
\affiliation{CAS Center for Excellence in Quantum Information and Quantum Physics, University of Science and Technology of China, Hefei 230026, China}

\author{Yi-Zheng Zhen}
\affiliation{Hefei National Research Center for Physical Sciences at the Microscale and School of Physical Sciences, University of Science and Technology of China, Hefei 230026, China}
\affiliation{CAS Center for Excellence in Quantum Information and Quantum Physics, University of Science and Technology of China, Hefei 230026, China}

\author{Yu-Xiang Yang}
\affiliation{National Laboratory of Solid State Microstructures, School of Physics, Nanjing University, Nanjing 210093, China}
\affiliation{Collaborative Innovation Center of Advanced Microstructures, Nanjing 210093, China}

\author{Zi-Mo Cheng}
\affiliation{National Laboratory of Solid State Microstructures, School of Physics, Nanjing University, Nanjing 210093, China}
\affiliation{Collaborative Innovation Center of Advanced Microstructures, Nanjing 210093, China}

\author{Zhi-Cheng Ren}
\affiliation{National Laboratory of Solid State Microstructures, School of Physics, Nanjing University, Nanjing 210093, China}
\affiliation{Collaborative Innovation Center of Advanced Microstructures, Nanjing 210093, China}

\author{Kai Chen}
\email{Corresponding author.\\kaichen@ustc.edu.cn}
\affiliation{Hefei National Research Center for Physical Sciences at the Microscale and School of Physical Sciences, University of Science and Technology of China, Hefei 230026, China}
\affiliation{CAS Center for Excellence in Quantum Information and Quantum Physics, University of Science and Technology of China, Hefei 230026, China}

\author{Xi-Lin Wang}
\email{Corresponding author.\\xilinwang@nju.edu.cn}
\affiliation{National Laboratory of Solid State Microstructures, School of Physics, Nanjing University, Nanjing 210093, China}
\affiliation{Collaborative Innovation Center of Advanced Microstructures, Nanjing 210093, China}

\author{Hui-Tian Wang}
\email{Corresponding author.\\htwang@nju.edu.cn}
\affiliation{National Laboratory of Solid State Microstructures, School of Physics, Nanjing University, Nanjing 210093, China}
\affiliation{Collaborative Innovation Center of Advanced Microstructures, Nanjing 210093, China}

\maketitle

\section{Measurement bases for the optimal quantum strategy}

\begin{table}
\begin{tabular}{cc|c|c|c|}
\multirow{1}{*}{} & \multicolumn{1}{c}{} & \multicolumn{3}{c}{$\overbrace{\hspace{3.2cm}}^y$}\tabularnewline[-.2cm]
& \multicolumn{1}{c}{\multirow{1}{*}{}} & \multicolumn{1}{c}{0} & \multicolumn{1}{c}{1} & \multicolumn{1}{c}{2}\tabularnewline
\cline{3-5} \cline{4-5} \cline{5-5}
\multirow{9}{*}{$x\left\{\rule{0pt}{40pt}\right.$}\hspace{-.2cm} &&&&\tabularnewline
 & 0 & $I\otimes Z$ & $Z\otimes I$ & $Z\otimes Z$\tabularnewline
&&&&\tabularnewline
\cline{3-5} \cline{4-5} \cline{5-5}
&&&&\tabularnewline
& 1 & $X\otimes I$ & $I\otimes X$ & $X\otimes X
 $\tabularnewline
&&&&\tabularnewline
\cline{3-5} \cline{4-5} \cline{5-5}
&&&&\tabularnewline
 & 2 & $-X\otimes Z$ & $-Z\otimes X$ & $Y\otimes Y$\tabularnewline
&&&&\tabularnewline
\cline{3-5} \cline{4-5} \cline{5-5}
\end{tabular}
\caption{\label{tab:quantstrategy}%
{\em Optimal quantum strategy}. The $X,Y,Z$ are three Pauli matrices.
When receiving queries $x$ and $y$, Alice and Bob select the $x$th row ($y$th column) of observables to measure their systems.
They win all queries with probability $1$.}
%\end{subtable}
%\end{minipage}
\end{table}

Quantum strategies can successfully fill every entry of the square with probability 1.
Suppose that in each round two maximally entangled state in the form of
\begin{equation}\label{equ}
\left|\Psi\right\rangle _{A_{1}A_{2}B_{1}B_{2}}=\frac{\left|00+11\right\rangle_{A_{1}B_{1}}}{\sqrt{2}} \otimes\frac{\left|00+11\right\rangle_{A_{2}B_{2}}}{\sqrt{2}}
\end{equation}
are distributed to Alice and Bob, where Alice has systems $A_{1}A_{2}$ and Bob has $B_{1}B_{2}$.
After the referee sends them queries $x$ and $y$, respectively, they independently perform measurements $M_{A}^{\left(x,i\right)}$ and $M_{B}^{\left(y,j\right)}$ and obtain answers $a^x_i$ and $b^y_j$.
The measurement strategy is given in Table.~\ref{tab:quantstrategy}, the same with Table.II in the main text.
In this table, the measurements in each row or in each column are pairwise commutative, such that they can be simultaneously measured.
It can be verified that for all $x$ and $y$, the product of Alice's output is always $+1$ becasue $\prod_{i}M_A^{(x,i)}=I$, while the product of Bob's output is always $-1$ because $\prod_{j}M_B^{(y,j)}=-I$.
For the game, Alice and Bob will always win each entry since $\left\langle \Psi\right|M_A^{(x,y)}\otimes M_B^{(y,x)}\left|\Psi\right\rangle =1$ $\forall x,y$, which means that they have same value $a^x_y=b^y_x$ such that quantum pseudotelepathy is exhibited.
It can be verified that for all $x$ and $y$, the product of Alice's output is always $+1$ because $\prod_{i}M_A^{(x,i)}=I$, while the product of Bob's output is always $-1$ because $\prod_{j}M_B^{(y,j)}=-I$.
For the game, Alice and Bob will always win each entry since $\left\langle \Psi\right|M_A^{(x,y)}\otimes M_B^{(y,x)}\left|\Psi\right\rangle =1$ $\forall x,y$, which means that they have same value $a^x_y=b^y_x$ such that quantum pseudotelepathy is exhibited.
% As a comparison, recall that for the usual nonlocal game (e.g., CHSH game), quantum strategies cannot win each query with probability 1 (CHSH game with a win probability of 85\% in quantum strategies).
% In this sense, quantum pseudotelepathy is a stronger form of nonlocality.

Here, we provide the bases to perform the compatible measurements as shown in Tables S2 and S3 for Alice and Bob, respectively.
%To simultaneously realize the measurement in each row or in each column as shown in Table.~\ref{tab:quantstrategy}, we choose the measurement basis as shown in Tables S2 and S3 for Alice and Bob, respectively, where
In both tables, we have let $\left|\pm\right\rangle =\left|0\pm 1\right\rangle /\sqrt{2}$, $\left|\alpha_{\pm\pm}\right\rangle $ be four orthogonal graph states
\begin{align}
\left|\alpha_{++}\right\rangle  & =\frac{\left|00+01+10-11\right\rangle }{2},\\
\left|\alpha_{+-}\right\rangle  & =\frac{\left|00-01+10+11\right\rangle }{2},\\
\left|\alpha_{-+}\right\rangle  & =\frac{\left|00+01-10+11\right\rangle }{2},\\
\left|\alpha_{--}\right\rangle  & =\frac{\left|00-01-10-11\right\rangle }{2},
\end{align}
and $\left|\beta_{\pm\pm}\right\rangle $ be four orthogonal Bell states
and can be measured by Bell state measurement. The $\left|\beta_{\pm\pm}\right\rangle $
have the form of
\begin{align}
\left|\beta_{++}\right\rangle  & =\frac{\left|00+11\right\rangle }{\sqrt{2}},\\
\left|\beta_{+-}\right\rangle  & =\frac{\left|01+10\right\rangle }{\sqrt{2}},\\
\left|\beta_{-+}\right\rangle  & =\frac{\left|00-11\right\rangle }{\sqrt{2}},\\
\left|\beta_{--}\right\rangle  & =\frac{\left|01-10\right\rangle }{\sqrt{2}}.
\end{align}
Upon the above notions, it can be verified that $a_{y}^{x}=b_{x}^{y}$
for any $x,y$ as $\left\langle \Psi\right|M_A^{(x,y)}\otimes M_B^{(y,x)}\left|\Psi\right\rangle =1$.

\renewcommand{\arraystretch}{.3}
\begin{center}
\begin{tabular}{>{\centering}m{5em}|>{\centering}m{13em}|>{\centering}m{5em}|>{\centering}m{10em}}
\multicolumn{4}{c}{TABLE S2: Alice's strategy}\tabularnewline[1em]
\hline \\[.3pt]
Query & Measurement $M^x$ & Result & Replied answer\tabularnewline[.2em]
\hline
\hline
\multirow{12}{*}{$x=0$} & \multirow{12}{*}{$Z\otimes Z$} &  & \tabularnewline
 &  & $\left|00\right\rangle $ & $\left[\begin{array}{ccc}
+1, & +1, & +1\end{array}\right]$\tabularnewline
 &  &  & \tabularnewline
\cline{3-4} \cline{4-4}
 &  &  & \tabularnewline
 &  & $\left|01\right\rangle $ & $\left[\begin{array}{ccc}
-1, & +1, & -1\end{array}\right]$\tabularnewline
 &  &  & \tabularnewline
\cline{3-4} \cline{4-4}
 &  &  & \tabularnewline
 &  & $\left|10\right\rangle $ & $\left[\begin{array}{ccc}
+1, & -1, & -1\end{array}\right]$\tabularnewline
 &  &  & \tabularnewline
\cline{3-4} \cline{4-4}
 &  &  & \tabularnewline
 &  & $\left|11\right\rangle $ & $\left[\begin{array}{ccc}
-1, & -1, & +1\end{array}\right]$\tabularnewline
 &  &  & \tabularnewline
\hline
\multirow{12}{*}{$x=1$} & \multirow{12}{*}{$X\otimes X$} &  & \tabularnewline
 &  & $\left|++\right\rangle $ & $\left[\begin{array}{ccc}
+1, & +1, & +1\end{array}\right]$\tabularnewline
 &  &  & \tabularnewline
\cline{3-4} \cline{4-4}
 &  &  & \tabularnewline
 &  & $\left|+-\right\rangle $ & $\left[\begin{array}{ccc}
+1, & -1, & -1\end{array}\right]$\tabularnewline
 &  &  & \tabularnewline
\cline{3-4} \cline{4-4}
 &  &  & \tabularnewline
 &  & $\left|-+\right\rangle $ & $\left[\begin{array}{ccc}
-1, & +1, & -1\end{array}\right]$\tabularnewline
 &  &  & \tabularnewline
\cline{3-4} \cline{4-4}
 &  &  & \tabularnewline
 &  & $\left|--\right\rangle $ & $\left[\begin{array}{ccc}
-1, & -1, & +1\end{array}\right]$\tabularnewline
 &  &  & \tabularnewline
\hline
\multirow{12}{*}{$x=2$} & \multirow{12}{*}{Graph state measurement} &  & \tabularnewline
 &  & $\left|\alpha_{++}\right\rangle $ & $\left[\begin{array}{ccc}
-1, & -1, & +1\end{array}\right]$\tabularnewline
 &  &  & \tabularnewline
\cline{3-4} \cline{4-4}
 &  &  & \tabularnewline
 &  & $\left|\alpha_{+-}\right\rangle $ & $\left[\begin{array}{ccc}
-1, & +1, & -1\end{array}\right]$\tabularnewline
 &  &  & \tabularnewline
\cline{3-4} \cline{4-4}
 &  &  & \tabularnewline
 &  & $\left|\alpha_{-+}\right\rangle $ & $\left[\begin{array}{ccc}
+1, & -1, & -1\end{array}\right]$\tabularnewline
 &  &  & \tabularnewline
\cline{3-4} \cline{4-4}
 &  &  & \tabularnewline
 &  & $\left|\alpha_{--}\right\rangle $ & $\left[\begin{array}{ccc}
+1, & +1, & +1\end{array}\right]$\tabularnewline
 &  &  & \tabularnewline
\hline
\end{tabular}
\par\end{center}
\renewcommand{\arraystretch}{1}
\begin{center}
\begin{tabular}{c||c|c|c|c|c|c|c|c|c|c|c|c}

\multicolumn{13}{c}{TABLE S3: Bob's strategy}\tabularnewline
\hline
Query & \multicolumn{4}{c|}{$y=0$} & \multicolumn{4}{c|}{$y=1$} & \multicolumn{4}{c}{$y=2$}\tabularnewline
\hline
Measurement $M^y$ & \multicolumn{4}{c|}{$-X\otimes Z$} & \multicolumn{4}{c|}{$-Z\otimes X$} & \multicolumn{4}{c}{Bell state measurement}\tabularnewline
\hline
Result & $\left|+0\right\rangle $ & $\left|+1\right\rangle $ & $\left|-0\right\rangle $ & $\left|-1\right\rangle $ & $\left|0+\right\rangle $ & $\left|0-\right\rangle $ & $\left|1+\right\rangle $ & $\left|1-\right\rangle $ & $\left|\beta_{++}\right\rangle $ & $\left|\beta_{+-}\right\rangle $ & $\left|\beta_{-+}\right\rangle $ & $\left|\beta_{--}\right\rangle $\tabularnewline
\hline
&&&&&&&&&&&&\tabularnewline
Replied answer & $\left[\begin{array}{c}
+1\\
+1\\
-1
\end{array}\right]$ & $\left[\begin{array}{c}
-1\\
+1\\
+1
\end{array}\right]$ & $\left[\begin{array}{c}
+1\\
-1\\
+1
\end{array}\right]$ & $\left[\begin{array}{c}
-1\\
-1\\
-1
\end{array}\right]$ & $\left[\begin{array}{c}
+1\\
+1\\
-1
\end{array}\right]$ & $\left[\begin{array}{c}
+1\\
-1\\
+1
\end{array}\right]$ & $\left[\begin{array}{c}
-1\\
+1\\
+1
\end{array}\right]$ & $\left[\begin{array}{c}
-1\\
-1\\
-1
\end{array}\right]$ & $\left[\begin{array}{c}
+1\\
+1\\
-1
\end{array}\right]$ & $\left[\begin{array}{c}
-1\\
+1\\
+1
\end{array}\right]$ & $\left[\begin{array}{c}
+1\\
-1\\
+1
\end{array}\right]$ & $\left[\begin{array}{c}
-1\\
-1\\
-1
\end{array}\right]$\tabularnewline
&&&&&&&&&&&&\tabularnewline
\hline
\end{tabular}
\par\end{center}

\section{The measurement scheme}

The crucial part for our experiment is the generation of hyperentangled state.
Note that, if one prepares the entangled states individually, say, two entangled photon pairs, and perform four-photon coincidence in the measurement, not only the setup can be complicated, but also the detection efficiency will significantly drop down.
Instead, we seek a hyperentangled photon pair and encode two entangled states into 2 DoFs, i.e.,  polarization and OAM\@.
We use $|H\rangle$, $|V\rangle$, $|r\rangle$, and $|l\rangle$ to represent the horizontal polarization, vertical polarization, right-handed OAM of $+\hbar$, and left-handed OAM of $-\hbar$, respectively.
The prepared state is in the form of
\begin{equation}\label{eq:exp-state}
\left|\Psi\right\rangle = \left|\psi\right\rangle _{{\rm pol}}\ensuremath{\otimes\left|\psi\right\rangle _{{\rm oam}}=\frac{\left|HH\right\rangle +\left|VV\right\rangle }{\sqrt{2}}\otimes\frac{\left|rl\right\rangle +\left|lr\right\rangle }{\sqrt{2}}.}
\end{equation}

To make the hyperentangled two photon state in Eq. (\ref{eq:exp-state}) have the same form in Eq.~\eqref{equ}, we encode the polarization qubit  $\left|H/V\right\rangle $ as  $\left|0/1\right\rangle $ for both photons, and encode the OAM qubit $\left|r/l\right\rangle $ as  $\left|0/1\right\rangle $ for Alice, while, encode the OAM qubit $\left|l/r\right\rangle $ as  $\left|0/1\right\rangle $ for Bob.
Below are the details about Alice and Bob's measurement apparatuses.

\subsection{Alice's measurement apparatus}

For $ZZ$ basis:
by passing through the PIP(CNOT) gate, the four basis are actually remains unchanged
\begin{align}
	&|00\rangle=|H,r\rangle \rightarrow |H,r\rangle,~~|01\rangle=|H,l\rangle\rightarrow |H,l\rangle,\\
	&|10\rangle=|V,r\rangle \rightarrow |V,l\rangle,~~|11\rangle=|V,l\rangle\rightarrow |V,r\rangle.
\end{align}

For $XX$ basis:
before enter the CNOT gate, the $XX$ basis is first transformed to $ZX$ basis by a HWP(1) at $22.5^{\circ}$, Passing through CNOT gate, the output from the interferometer reads
\begin{align}
	&|++\rangle=(|H+V)\rangle(|r\rangle+|l\rangle)/2 \stackrel{HWP1}{\longrightarrow} |H\rangle(|r\rangle+|l\rangle)\stackrel{CNOT}{\longrightarrow} |H\rangle(|r\rangle+|l\rangle),\\
	&|+-\rangle=(|H-V\rangle)(|r\rangle-|l\rangle)/2 \stackrel{HWP1}{\longrightarrow} |H\rangle(|r\rangle-|l\rangle)\stackrel{CNOT}{\longrightarrow} |H\rangle(|r\rangle-|l\rangle),\\
	&|-+\rangle=(|H+V\rangle)(|r\rangle+|l\rangle)/2 \stackrel{HWP1}{\longrightarrow} |V\rangle(|r\rangle+|l\rangle)\stackrel{CNOT}{\longrightarrow} |V\rangle(|l\rangle+|r\rangle),\\
	&|--\rangle=(|H-V\rangle)(|r\rangle-|l\rangle)/2 \stackrel{HWP1}{\longrightarrow} |V\rangle(|r\rangle-|l\rangle)\stackrel{CNOT}{\longrightarrow} |V\rangle(|l\rangle-|r\rangle).
\end{align}

For Graph state basis: the HWP(1$\sim2$) should be set to $22.5^\circ$, the basis changes as follow:
\begin{align}
	\left|\alpha_{++}\right\rangle  & =(|Hr\rangle+|Hl\rangle+|Vr\rangle-|Vl\rangle)/2 \stackrel{HWP1}{\longrightarrow} |Ar\rangle+|Al\rangle+|Dr\rangle-|Dl\rangle \stackrel{CNOT}{\longrightarrow} |Ar\rangle \stackrel{HWP2}{\longrightarrow} |Hr\rangle,\\
	\left|\alpha_{+-}\right\rangle  & =(|Hr\rangle-|Hl\rangle+|Vr\rangle+|Vl\rangle)/2 \stackrel{HWP1}{\longrightarrow} |Ar\rangle-|Al\rangle+|Dr\rangle+|Dl\rangle \stackrel{CNOT}{\longrightarrow} |Dr\rangle \stackrel{HWP2}{\longrightarrow} |Vr\rangle,\\
	\left|\alpha_{-+}\right\rangle  & =(|Hr\rangle+|Hl\rangle-|Vr\rangle+|Vl\rangle)/2 \stackrel{HWP1}{\longrightarrow} |Ar\rangle+|Al\rangle-|Dr\rangle+|Dl\rangle \stackrel{CNOT}{\longrightarrow} |Al\rangle \stackrel{HWP2}{\longrightarrow} |Hl\rangle,\\
	\left|\alpha_{--}\right\rangle  & =(|Hr\rangle-|Hl\rangle-|Vr\rangle- |Vl\rangle)/2 \stackrel{HWP1}{\longrightarrow} |Ar\rangle-|Al\rangle-|Dr\rangle-|Dl\rangle \stackrel{CNOT}{\longrightarrow} |Dl\rangle \stackrel{HWP2}{\longrightarrow} |Vl\rangle.
\end{align}

\begin{figure}[!htp]
\setlength{\abovecaptionskip}{-2 mm}
\setlength{\belowcaptionskip}{-5 mm}
\centering
\includegraphics[width=0.95\linewidth]{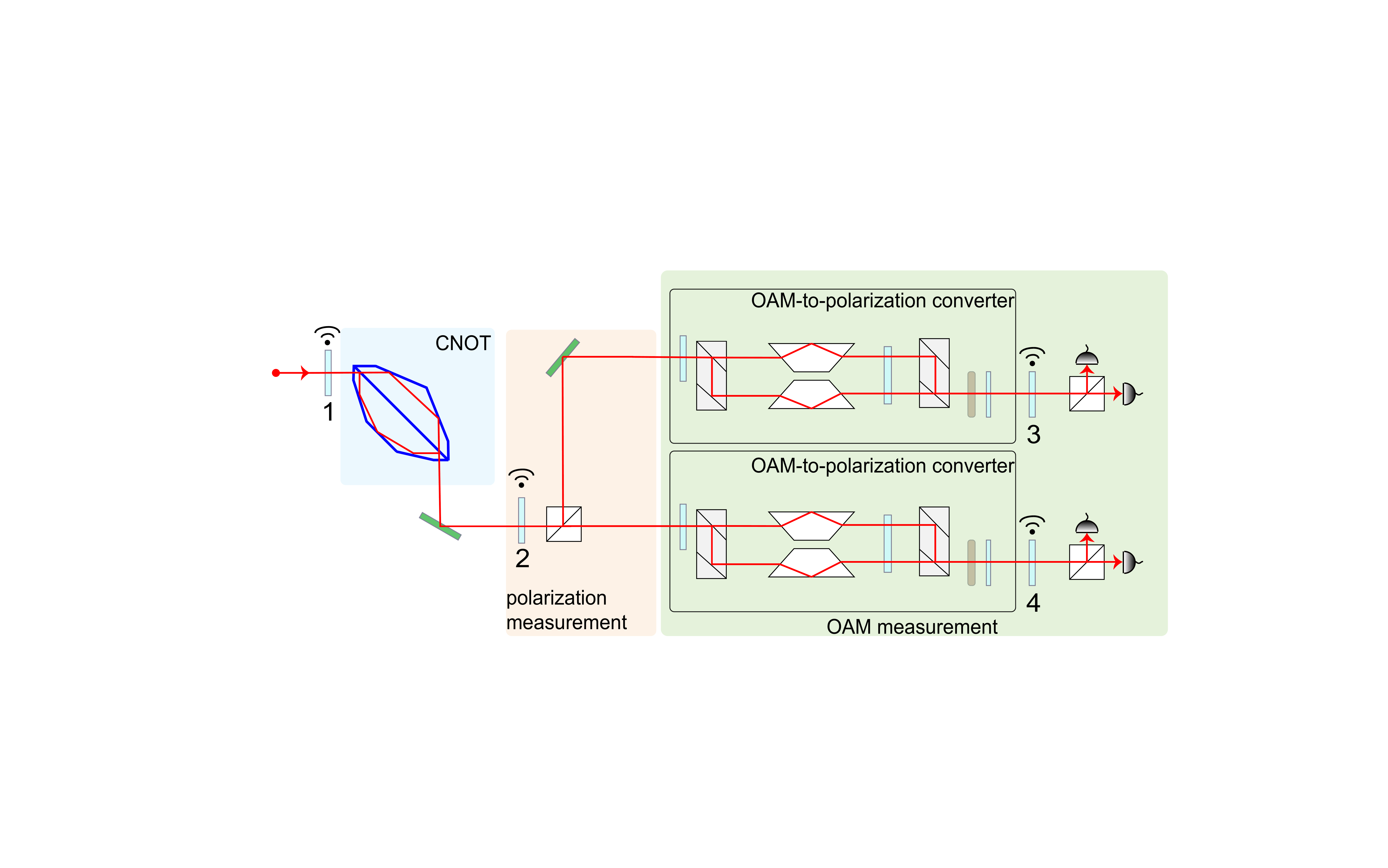}\vspace{3mm}
\caption{Alice/Bob's measurement apparatus}\vspace{5mm}
\end{figure}

With above deterministic transformation, the photon in all the Alice's measurement basis states have been transferred to product states between polarization and OAM, then we can read the information in these two degrees one by one following polarization measurement and OAM measurement~\cite{wang201818}. %[X.-L. Wang \textit{et al.}, Phys. Rev. Lett. \textbf{120}, 260502 (2018)].

\subsection{Bob's measurement apparatus}
For $XZ$ basis: the HWP1 at $22.5^\circ$ transform the polarization to the $H/V$ then remains unchanged after passing through the CNOT gate.
\begin{align}
	|+0\rangle &=|(H+V)l\rangle/\sqrt{2} \stackrel{HWP1}{\longrightarrow} |Hl\rangle \stackrel{CNOT}{\longrightarrow} |Hl\rangle,\\
	|+1\rangle &=|(H+V)r\rangle/\sqrt{2} \stackrel{HWP1}{\longrightarrow} |Hr\rangle \stackrel{CNOT}{\longrightarrow} |Hr\rangle, \\
	|-0\rangle &=|(H-V)r\rangle/\sqrt{2} \stackrel{HWP1}{\longrightarrow} |Vr\rangle \stackrel{CNOT}{\longrightarrow} |Vl\rangle,\\
	|-1\rangle &=|(H-V)l\rangle/\sqrt{2} \stackrel{HWP1}{\longrightarrow} |Vl\rangle \stackrel{CNOT}{\longrightarrow} |Vr\rangle.
\end{align}

For $ZX$ basis: the HWP1 amd HWP2 can be set to $0^\circ$, the basis transformation is as follow:
\begin{align}
	|0+\rangle &=|H\rangle(|l\rangle+|r\rangle) /\sqrt{2}\stackrel{CNOT}{\longrightarrow} |H\rangle(|l\rangle+|r\rangle),\\
	|0-\rangle &=|H\rangle(|l\rangle-|r\rangle)/\sqrt{2} \stackrel{CNOT}{\longrightarrow} |H\rangle(|l\rangle-|r\rangle),\\
	|1+\rangle &=|V\rangle(|l\rangle+|r\rangle)/\sqrt{2} \stackrel{CNOT}{\longrightarrow} |V\rangle(|r\rangle+|l\rangle),\\
	|1-\rangle &=|V\rangle(|l\rangle-|r\rangle) /\sqrt{2}\stackrel{CNOT}{\longrightarrow} |V\rangle(|r\rangle-|l\rangle).
\end{align}

For Bell basis: the HWP1 and HWP2 should set $0^\circ$ and $22.5^\circ$, respectively.
\begin{align}
|\beta_{++}\rangle   &=(\left|Hl\rangle+|Vr\right\rangle)/\sqrt{2} \stackrel{CNOT}{\longrightarrow} |Al\rangle \stackrel{HWP2}{\longrightarrow} |Hl\rangle,\\
|\beta_{+-}\rangle   &=(\left|Hr\rangle+|Vl\right\rangle)/\sqrt{2} \stackrel{CNOT}{\longrightarrow} |Ar\rangle \stackrel{HWP2}{\longrightarrow} |Hr\rangle,\\
|\beta_{-+}\rangle   &=(\left|Hl\rangle-|Vr\right\rangle)/\sqrt{2} \stackrel{CNOT}{\longrightarrow} |Dl\rangle \stackrel{HWP2}{\longrightarrow} |Vl\rangle,\\
|\beta_{--}\rangle   &=(\left|Hr\rangle-|Vl\right\rangle)/\sqrt{2} \stackrel{CNOT}{\longrightarrow} |Dr\rangle \stackrel{HWP2}{\longrightarrow} |Vr\rangle.
\end{align}

Similarly, when the photon state in two degrees of freedom is transferred to a product state, we could measure it first in polarization and then in OAM\@.

In conclusion, only four HWPs(1$\sim$4) that either $0^\circ$ or $22.5^\circ$ are needed to meeting the experimental requirements. The angle of the HWP are decided by the measurement basis, the correspondence between the basis and the angle of HWPs(1$\sim$4) are listed in Table S4.

\begin{center}
TABLE S4. Correspondence between basis and angles of HWPs
\begin{tabular}{|c|c|c|c|c|c|}
	\hline
	Player & Measurement basis & HWP1  & HWP2 & HWP3 & HWP4 \\
	\hline
	\multirow{3}{*}{Alice} & $ZZ$ & $0^\circ$ & $0^\circ$ & $0^\circ$ & $0^\circ$ \\
	\cline{2-6}
	& $XX$ & $22.5^\circ$ & $0^\circ$ & $22.5^\circ$ & $22.5^\circ$ \\
	\cline{2-6}
	& $\alpha_{\pm\pm}$ & $22.5^\circ$ & $22.5^\circ$ & $0^\circ$ & $0^\circ$ \\
	\hline
	\multirow{3}{*}{Bob} & $XZ$ & $22.5^\circ$ & $0^\circ$ & $0^\circ$ & $0^\circ$ \\
	\cline{2-6}
	& $ZX$ & $0^\circ$ & $0^\circ$ & $22.5^\circ$ & $22.5^\circ$ \\
	\cline{2-6}
	& $\beta_{\pm\pm}$ & $0^\circ$ & $22.5^\circ$ & $0^\circ$ & $0^\circ$ \\
	\hline
\end{tabular}
\end{center}

\section{Minimal detection efficiency to close the detection loophole}

Here, we derive the minimal detection efficiency required to close the detection loophole.
Since this problem is related to the precise experimental setup, we will focus on the simple case of hyperentangled perfect source.
For convenience, the transmission loss of the photons are attributed into the inefficiency of detectors.
Suppose that all detectors have the same detection efficiency $\eta$ and the dark counts of detectors are zeros.
When using hyperentangled photon pairs, there is only one photon in Alice, Bob's lab, respectively.
Then, the $\eta$ is in fact the probability that Alice (or Bob) obtains a click in her (or his) detector.
When the photon is lost, Alice or Bob can only reply according to some preshared randomness.
Assume that the unbiased optimal classical strategy is adopted if the detector dose not click.
We consider the average winning probability of all query pairs, i.e.
\begin{equation}
I_{{\rm exp}}=\eta^{2}I_{Q}+\left(1-\eta\right)^{2}I_{C}+2\eta\left(1-\eta\right)I_{QC}.
\end{equation}
Here, with probability $\eta^{2}$, both photons arrive at Alice's and Bob's detectors, and the game is won with value $I_{Q}=1$.
With probability $\left(1-\eta\right)^{2}$, both photons are lost and Alice and Bob win the game with value $I_{C}=8/9$ according to the optimal classical strategy.
With probability $2\eta\left(1-\eta\right)$, one photon is lost and the other photon generates $\pm1$ results uniformly.
In this case, one can compute the game score as $I_{QC}=1/2$ since the lost photon case is replaced by a classical strategy.
One can compute the average winning probability as
\begin{align}
I_{{\rm exp}} & =\eta^{2}I_{Q}+\left(1-\eta\right)^{2}I_{C}+2\eta\left(1-\eta\right)I_{QC}\\
 & =\left(I_{Q}+I_{C}-2I_{QC}\right)\eta^{2}+\left(-2I_{C}+2I_{QC}\right)\eta+I_{C}\\
 & =\frac{8}{9}\eta^{2}-\frac{7}{9}\eta+\frac{8}{9}.
\end{align}
To win the game better than the classical strategy, one needs $I_{{\rm exp}}>8/9$
which yields
\begin{equation}
\eta>\frac{7}{8}=87.5\%.
\end{equation}

\bibliography{Ref_MagicSquare}